# Laser-assisted Cellular Electrophysiology Measurement System

A. A. Seymen, E. Özgür, Z. Soran-Erdem, B. Ortaç

*Abstract*—Patch-clamp technique is the gold standard for cellular electrophysiological measurements, which is capable of measuring single ion transport events across the cell membrane. However, the measurement possesses significant complexity, and it requires a high level of expertise, while its experimental throughput is nevertheless considerably low. Here, we suggest and experimentally demonstrate a laser-assisted method for performing cellular electrophysiological measurements. Femtosecond laser pulses, coupled to an optical microscope, are used to form a sub-micrometer hole on a thin polymer membrane separating two electrodes, where a nearby cell is subsequently placed onto the hole by negative pressure. Afterwards, the cell is punctured using subsequent laser exposure, revealing the cell membrane over the hole for electrophysiological recording. This system could be used to increase the output amount of the electrophysiological measurements substantially.

## I. INTRODUCTION

BESIDES isolating the cellular metabolism from the exterior, cell membranes gather intracellular components such as organelles and various metabolites [1]. Also, they host membrane proteins, which correspond to 20-30% of genes expressed in a cell [2] and are the targets for 50% of pharmaceutical drugs [3]. Although the main building block of cellular membranes are amphipathic phospholipids, which spontaneously form lipid bilayers under physiological conditions [4], membrane proteins have an important role determining the membrane, and therefore cellular characteristics, by controlling the uptake and excretion of materials [5], transducing signals from the extracellular environment [6], communicating with other cells [7], and anchoring cell to the extracellular matrix [8]. Thus, it is not surprising that the investigation of cellular membranes is a major field of scientific research.

Among the integral membrane proteins, ion channels are particularly important because of their role in maintaining the cellular homeostasis by acting like molecular switches operating under influences such as electrical potential [9], chemical mediators [10], mechanical [11], thermal [12], and even optical stimuli [13]. Nonetheless, their function merely depends on basic physical laws of nature [14, 15]. Thus, ion channel research forms a basis for a wide span of disciplines investigating the physiology and pathology of the cells. The main goals of electrophysiology research are anticipating the reactions of ion channels under the influence of different stimuli, and eventually controlling them using various methods. These efforts aim not only deciphering the underlying mechanisms of their operation, but also understanding and interfering with related pathological conditions, which are essential in discovering and developing novel drugs. Considering the drawbacks of isolation problems and obtaining same response from all isolated cells, H9c2 is a significant cell line source to investigate ion channels by electrophysiology. H9c2 cells are clonal myogenic cell line derived from embryonic rat ventricles which initially had been thought as myocytic, but later they were realized that they possess both skeletal and cardiac muscle features [16]. In a study published by Hescheler et al., researchers demonstrated that these cells preserve several electrical and hormonal signal pathways similar to adult cardiac cells [17]. After this founding, H9c2 cells became center of attraction because of their L-type channels and voltage-dependent nonspecific channels, especially for electrophysiology studies [18].

To date, patch-clamp recording is considered as the gold standard for measuring the ionic currents through ion channels inside and outside the cells. This technique, which inventors are Nobel laureates, enables direct observation of ionic fluxes [19] by using finely tapered glass capillaries for electrically insulating electrodes on reciprocal sides of ion channels that seal the cell membrane by suction. Using this technique, ion channels could be interrogated either as constituents of a small membrane patch sealed on tip of the cell membrane, to observe individual opening and closing of these tiny gates, or measuring the currents passing through the whole membrane. The power of this technique resides in the fact that one of the electrodes is within the capillary, while the other is in the bath solution. Therefore, as long as the fine tip of the capillary is sealed onto the cell membrane, using the electrostatic interactions among the thermally tapered glass and the phospholipids, the membrane patch, or the whole cell, acts

This study is partially supported by the Ministry of Science, Industry, and Technology of Turkey. (Corresponding authors: E. Ozgur and B. Ortaç)

A. A. Seymen is with the E-A Teknoloji LLC, 06800 Ankara, Turkey; Dept. Physiology, Erciyes University, 38039 Kayseri, Turkey; and UNAM - National Nanotechnology Research Center and Institute of Materials Science and Nanotechnology, Bilkent University, 06800 Ankara, Turkey (e-mail: aytac@eateknoloji.com).

E. Ozgur is with the E-A Teknoloji LLC, 06800 Ankara, Turkey (e-mail: erol@eateknoloji.com). Current Address: Wyant College of Optical Sciences, University of Arizona, Tucson, AZ, USA 85721 (erol@optics.arizona.edu)

Z. Soran-Erdem was with UNAM - National Nanotechnology Research Center and Institute of Materials Science and Nanotechnology, Bilkent University, 06800 Ankara, Turkey. Present address is Abdullah Gul University Engineering Science, 38080 Kayseri, Turkey (e-mail: zeliha.soranerdem@agu.edu.tr).

B. Ortaç is with the UNAM - National Nanotechnology Research Center and Institute of Materials Science and Nanotechnology, Bilkent University, 06800 Ankara, Turkey (e-mail: ortac@unam.bilkent.edu.tr).

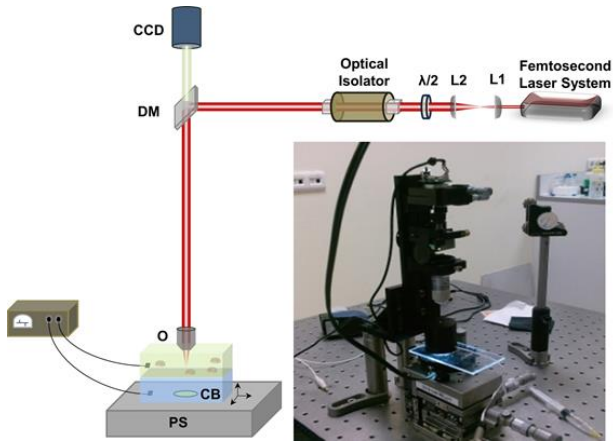

Fig. 1. Experimental setup. The Ti:Sapphire femtosecond laser is coupled to a custom-built microscope from its side-port. Laser illumination passes through a beam expander lens system, an optical isolator, and it is reflected towards the sample using a dichroic mirror (DM). It is focused using a 100X water immersion microscope objective (O), which is dipped into the bath solution. The image is collected using a CCD camera. In the cell bath (CB) the upper compartment has a physiological solution, and the other contains electrode solution. The electrodes are connected to a miniature patch-clamp amplifier. Positioning system (PS) is intended to be used to facilitate the delivery of cell to femtosecond laser pulse. The picture of the actual system is also given in the figure.

analogous to a RC circuit, which resistance is variable. This technique also enables administration of different chemicals including channel blockers, forming a perfect scheme for pharmacological studies.

Besides the unprecedented opportunities provided by patch-clamp measurement, this method is nevertheless one of the most cumbersome practices to apply experimentally. It requires a perfect isolation of the cells from the tissue, well defined parameters for capillary pulling, and a complex experimental set-up. Training for patch-clamp requires considerable time for the experimenter. The throughput of the experiment is generally low. Therefore, alternative techniques were investigated for simplifying the process. Planar patch-clamp systems, for instance, gained considerable reputation regarding their straightforward use. Instead of using capillaries, these systems utilize chips that have a micron-scale hole, where the cells are pulled by suction [20]. There are automated systems using many channels simultaneously, offering a tremendous increase in the output. However, these systems are blind; i.e., the users do not have control on the cell they are measuring. While that does not cause a major problem with cultured cells, isolated cells may have more cell to cell variation. Also, planar systems do not work well with large cells such as cardiac myocytes.

Here, we suggest a new technique for electrophysiological measurement of cells, laser-assisted cellular electrophysiology measurement (LACEM) system, using femtosecond laser pulses. Femtosecond lasers have previously been shown to create membrane depolarization [21], by forming sub-micron holes on cellular membrane, without harming the cells in general; yet, membrane potential and its alterations due to laser application were measured using a conventional patch-clamp system, rather than suggesting an alternative to the patch-clamp pipette. Besides other nanosurgery applications such as femtosecond laser-assisted transfection [22] or cellular sucrose deposition [23], which barely create an opportunity for large scale applications, laser induced membrane depolarization by itself also does not promise a wide applicability. In addition to membrane depolarization, ultrafast laser pulses are also shown to create micron sized holes on polymer films as well [24]. In this study, we demonstrate that combining these two separate works could be used to measure membrane currents throughout the cells.

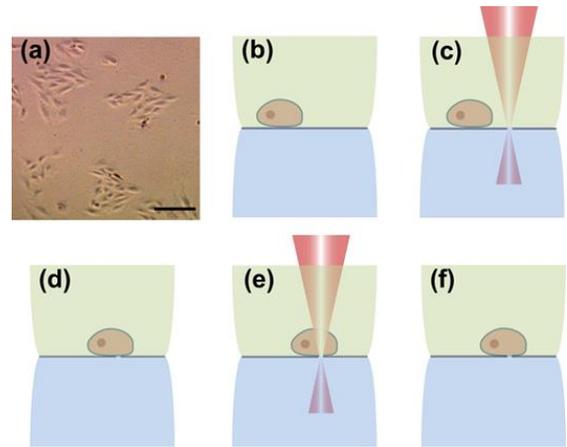

Fig. 2. Laser-assisted electrophysiological measurement. After the H9c2 cells grown in culture (a) (Scale bar = 200 microns) are harvested and placed into the upper compartment of the cell bath, and they settle down on the polymer film (b), a femtosecond laser pulse is used to punch the membrane separating the two baths, nearby a cell (c), and the cell is brought on top of the hole by suction (d). Afterwards, another laser pulse is used to punch a hole in the cell membrane (e) This causes the cell membrane to be exposed to the solution in the lower compartment, in a similar manner to a whole-cell patch clamp configuration (f).

## II. EXPERIMENTAL DESIGN

A cell bath is prepared by separating two cylinder shaped containers by attaching a thin Mylar film in between. Each cylinder has their own liquid inlet and outlet, and electrode insertion sealed with silicone. Here, the below part is the electrode solution, while the above part is the bath solution. H9c2 rat cardio myoblast cells were used in patch-clamp experiments because of their electrically excitable nature [18]. For this, a diluted cell culture suspension is dropped onto the Mylar film and the electrodes are connected to a Tecella Pico miniature patch-clamp amplifier. An upright microscope is used to observe the cells and experiment is conducted after the cells are settled down. The schematic representation of the experimental set-up is given in Fig. 1. Femtosecond laser pulses generated from a Ti:Sapphire laser is coupled to home-made microscope from its side part. The laser operates at 800 nm, which is a wavelength that has a minimal harm on living tissues [25]. The tail of the laser emission is visible to human eye, which is used for alignment of the laser beam.

## III. RESULTS AND DISCUSSION

At the beginning of the experiment, the electrode potential difference is a function of the resistivity and capacitance of the Mylar film. Then, the laser is first focused on the Mylar film just beneath, or beside the target cell, using a low laser power, and continuous wave emission. Afterwards, femtosecond pulses are sent over the sample and generate a slight suction, which causes opening of pores on both H9c2 cell and the polymer film. Here, the cell is spontaneously sealed on the newly formed hole. Schematic image of the cell on the set-up is presented in Fig. 2.

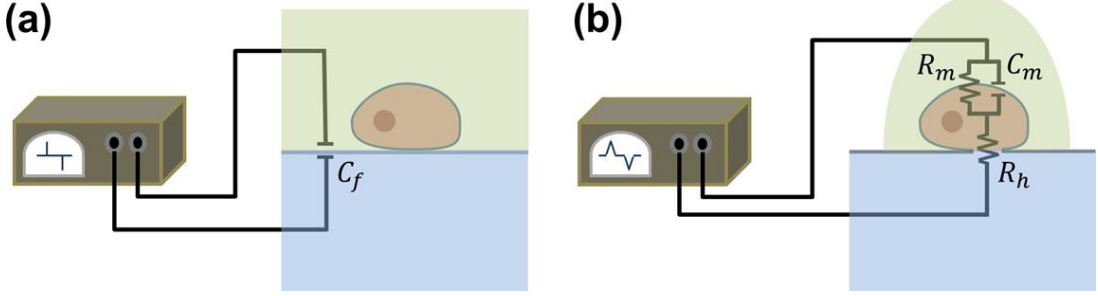

Fig. 3. Electrical analogue of the open cell using femtosecond laser. Before laser application, two electrodes are electrically isolated from each other, enabling observation of only short capacitive currents through the polymer membrane ($C_f$) (a). After the laser treatments and the liquid removal except the volume around the cell to reduce the $C_f$, the cell is now a RC circuit, in a similar fashion to patch clamp experiments. The major electrical components are hole and membrane resistances ($R_h$ and $R_m$, respectively), and the membrane capacitance ($C_m$), which is significantly lower than the $C_f$. Under these circumstances, membrane currents through the ion channels, caused by voltage pulse become obvious, which could be observed using a patch-clamp amplifier (b).

After the cell is sealed onto the hole formed on the Mylar film, and another hole is formed on the cell membrane, the cell becomes electrically analogous to a RC circuit, in a similar fashion to the patch-clamp using a micropipette (Fig. 3). Here, the examined electric current is formed by the ions passing through the ion channels of protein structure, embedded in the cell membrane. The membrane resistance varies as ion channels open and close. Also, there is a small capacitance of the cell membrane, and since the ion flow through the hole at the cell membrane will be compromised by the small dimension of it, the hole constitutes a second resistance, which has a fixed value. Before the hole on the Mylar film is formed, a voltage pulse of 20 ms (Fig. 4a) only creates an instantaneous capacitive current (Fig. 4b), while a hole opened on the Mylar film by the femtosecond laser (Fig. 4c) creates a constant current (Fig. 4d) with the same voltage pulse. After the suction is performed and the cell is placed on top of the hole (Fig. 4e), this current is subsided. Finally, the cell membrane is opened again with another femtosecond pulse, which enables the observation of currents through the voltage-dependent channels of H9c2 cells (Fig. 4f). Here, it is important to note that, after the final step of membrane opening, most of the liquid on the top compartment of the measurement chamber should be removed, since the capacitance is:

$$C_f = \epsilon_f \frac{A_f}{d_f} \quad (1)$$

where $C_f$ is the capacitance of the polymer film, $\epsilon_f$ is the electric permittivity of the Mylar, $A_f$ is the area of film between the compartments, and $d_f$ is the film thickness. The reduction of the area between the compartments minimizes the capacitive currents. On the other hand, removal of the liquid creates a lens effect, which might affect the focusing of the femtosecond laser. Thus, it is applied as the last step before the measurement.

## IV. CONCLUSION

In this study, we demonstrate a novel and convenient method for exploring the cellular membrane currents, by eliminating the need for using glass capillary tubes. This technique also enables selecting the cell that would be investigated. Different patch configurations, such as cell-attach, whole cell, inside out, and outside in could be mimicked by changing the initial hole size on the Mylar film, and constituents of the electrode and bath solutions. For instance, opening a hole on the Mylar membrane, and placing the cell on the hole by suction would be cell attach configuration. If the hole is large, it would be whole cell. If there would be a small hole, by changing the constituents of the electrode and bath solutions, inside out and outside in configurations would be attained.

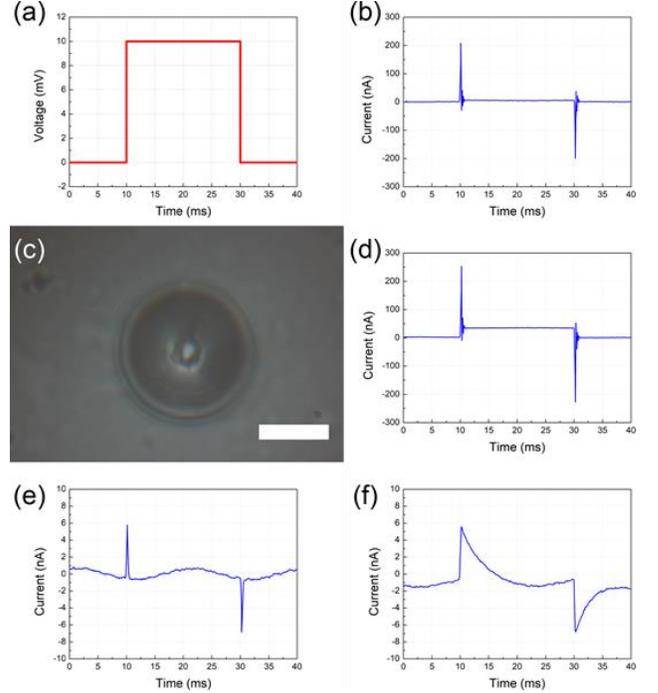

Fig. 4. Experimental demonstration of Laser-assisted cellular electrophysiology measurement. A voltage pulse of 20 ms is used to probe the membrane currents (a). Initially, there is only the capacitive current present (b). When a hole on the Mylar membrane is formed (c) (Scale bar = 1 micron), the corresponding current through the hole is observed (d). After the cell is sucked on top of the hole (e), a hole on the cell is formed, revealing electric currents through voltage gated channels of H2c9 cells (f). The liquid on the upper compartment, i.e., intracellular bath solution, is removed after sucking of the cell, refilled, and removed again, to reduce the effect of the capacitive currents.

LACEM system is flexible and suitable for being developed further into a widely applicable technique. Particularly the design of the cell bath chamber could also be further optimized without changing the operation principle. The only prerequisite is using an ultrashort pulse laser, operating at a wavelength that does not interact with a living tissue. Alternative femtosecond lasers, for instance [26], can replace Ti:Sapphire laser systems, which are bulky and expensive. The patch-clamp amplifier we used in this study could also be superseded by better equipment; yet, they would not be as cost-effective as described in this work. Our major aim was to demonstrate an example towards a facile and low-cost electrophysiological measurement system, which we believe that we provide sufficient experimental evidence.

Further studies are necessary to fully characterize our system, particularly using different types of cells under different protocols. On the other hand, this study is, to our knowledge, still the very first demonstration of a laser-assisted electrophysiological measurement system, providing clear insights regarding its applicability. It is among our future plans to continue investigation related to this study and explore the possibilities in terms of transforming this knowledge into a product that would be well-established and widely accessible.

As a conclusion, we have successfully developed a facile and novel system, using femtosecond laser nanosurgery, to measure the activity of ion channels, in a similar fashion to patch-clamp recording, the gold standard in electrophysiology, paving the way for higher throughput electrophysiological measurements.

ACKNOWLEDGMENT

The authors would like to thank to Prof. Ümit Bağrıaçık from Gazi University for providing H9c2 cell line and Nanobiotechnology Group at Bilkent University, UNAM for their support in cell culture maintenance. A.A.S. and E.Ö. declare financial interest regarding LACEM through E-A Teknoloji LLC.